\documentclass[referee]{raa} 
\usepackage{graphicx,times}
\usepackage{natbib}
\usepackage{amssymb,amsmath}
\bibpunct{(}{)}{;}{a}{}{,}

\usepackage[driverfallback=dvips]{hyperref}
\hypersetup{pdftitle = The title of my PDF, pdfauthor = My name, pdfsubject= The subject, pdfkeywords = keyword1 keyword2 keyword3} 
\hypersetup{colorlinks = true, linkcolor = green, anchorcolor = red, citecolor = blue, filecolor = red, pagecolor = red, urlcolor = red}

\begin{document}

   \title{\bf The role of mergers and gas accretion in black hole growth and galaxy evolution}

 \volnopage{ {\bf 20XX} Vol.\ {\bf X} No. {\bf XX}, 000--000}
   \setcounter{page}{1}

   \author{Tianchi Zhang\inst{1,2}, Qi Guo\inst{1,2}, Yan Qu\inst{1}, Liang Gao
      \inst{1,2,3}
   }

   \institute{Key Laboratory for Computational Astrophysics, National Astronomical Observatories, Chinese Academy of Sciences, Beijing 100012, China; {\it guoqi@nao.cas.cn}\\
        \and
             School of Astronomy and Space Science, University of Chinese Academy of Sciences, Beijing 100049, China\\
	\and
Institute of Computational Cosmology, Department of Physics, University of Durham, South Road, Durham DH1 3LE, UK\\
\vs \no
   {\small Received 20XX Month Day; accepted 20XX Month Day}
}

\abstract{We use a semi-analytic galaxy formation model to study the co-evolution of supermassive black holes (SMBHs) with their host galaxies. Although the coalescence of SMBHs is not important, the quasar-mode accretion induced by mergers plays a dominant role in the growth of SMBHs. Mergers play a more important role in the growth of SMBH host galaxies than in the SMBH growth. It is the combined contribution from quasar mode accretion and mergers to the SMBH growth and the combined contribution from starburst and mergers to their host galaxy growth that determine the observed scaling relation between the SMBH masses and their host galaxy masses. We also find that mergers are more important in the growth of SMBH host galaxies compared to normal galaxies which share the same stellar mass range as the SMBH host galaxies. 
}

   \authorrunning{T.C. Zhang et al. }            
   \titlerunning{Supermassive black hole assembly}  
   \maketitle

%
\section{Introduction}\label{sec_intro}

The co-evolution of the SMBHs and their host galaxies is one of the most important topics in modern cosmology. SMBH masses closely correlate with the bulge properties of their host galaxies, including the bulge velocity dispersion, K-band luminosity and stellar masses \citep[e.g.][]{Ferrarese2000, Ueda2003, Zheng2009,McConnell2013,Kormendy2013,Greene2016,Schutte2019}. Although SMBHs cannot be detected directly, mass accretion onto SMBHs can power exotic phenomena, the active galactic nuclei (AGNs) activities, which are usually very luminous in X-ray, radio and optical wavelengths \citep[see e.g.][etc]{Shankar2009b, Ueda2014, Buchner2015, Nicola2019, Thater2019, Reines2020}. The volume density of AGN peaks at 2$\sim$3 \citep{Ferrarese2000}, similar to that of the star formation rate density \citep{Hopkins2006} and to that of mergers, suggesting connections among SMBHs growth, star formation and mergers.  

SMBHs gain mass via accretion of gas and mergers. \cite{Croton2006} proposed that there are two modes of gas accretion in modern galaxy formation theories: radio mode and quasar mode. In the radio mode, SMBHs grow through accreting hot gas in a relatively quiescent way. AGN feedback in this mode is essential to prevent  gas cooling in the centers of clusters. The quasar mode accretion is triggered by merging of gas-rich galaxies. Violent processes during mergers reduce the angular momentum dramatically, leading to gas fueling into the center SMBH of the remnant galaxy. This can be an important mechanism to build up the correlation between the SMBH and their host galaxies, especially at high redshift when gas-rich mergers are more common. In hydrodynamical simulations, these two modes are usually separated according to the accretion rate-to-Eddington accretion rate ratio
\citep[e.g.][]{Dubois2014, Weinberger2018}.

SMBHs can also grow by coalescence of progenitor SMBHs. This is in line with the fact that at high redshifts extremely luminous AGNs are found to exclusively reside in disturbed systems \citep[e.g.][]{Fan2016}. Yet for non-extremely luminous AGNs and at lower redshifts, the connection between SMBH growth and galaxy mergers are still under debate. Some studies \citep[e.g.][]{Cotini2013,Satyapal2014,Ellison2015,Goulding2018} found that AGN activities are boosted in close galaxy pairs and disturbed galaxies. On the other hand, other studies \citep[e.g.][]{Gabor2009,Schawinski2011,Karouzos2014,Hewlett2017} found no connection between the AGN triggering/fueling and major mergers. \citet{Marian2019} claimed that even at the highest specific accretion rates and at the peak of cosmic AGN activity, the SMBH growth is not driven by major mergers. Discrepancies in the observational results could be attributed to multiple factors, including 
dust obscuration, morphology identification, sample variance, etc.

Simulations are compelling tools to understand the roles different mechanisms play in the growth of the SMBH for one can trace the growth over cosmic time. Some works found that mergers could drive the SMBH growth at high masses and at low redshifts. Assuming that only merger-related processes are relevant, \citet{Malbon2007} found that the gas accretion induced by mergers is more important for less massive SMBHs at high redshift, while coalescence of SMBHs drives the growth of massive SMBHs at low redshift. \citet{Marulli2007} claimed that the merger-driven scenario is not capable of reproducing the abundance of faint X-ray sources in the local Universe. Taking into account both radio mode and quasar mode accretion, \citet{Fanidakis2011} concluded that the SMBH growth is dominated by the quasar mode accretion at high redshifts, while at $z=0$ mergers and radio mode accretion become the dominant channels at high masses. \citet{Marshall2020} found that the quasar mode, especially the instability induced quasar mode accretion, dominates the SMBHs growth. \citet{Bonoli2009} inferred that accretion is very important for low mass SMBHs, and for high mass SMBHs at high redshifts. Similar results are also found in hydrodynamical simulations \citep{Dubois2014} and semi-analytical galaxy formation models \citep{Croton2006}. In IllustrisTNG simulations \cite{illTNG} and  \cite{Weinberger2018} found the instantaneous growth is dominated by mergers for massive SMBHs, yet by accretion for low-mass SMBHs. However, \citet{McAlpine2020} found in the EAGLE simulation \citep{eagle} that mergers do not contribute a significant fraction of the final mass of the SMBHs. \citet{Lapiner2020} studied the interaction between galaxy evolution and central SMBH growth in the NewHorizon cosmological simulation\citep{Dubois2020}. They found a golden halo mass below which the SMBH growth slows, while above which the SMBHs grow rapidly. Mergers play an important role in triggering the onset of the SMBH growths.
It is still under debate whether mergers could be the driving mechanism in the growth of the SMBHs, especially at high masses.

Corresponding to the growth of SMBHs, galaxies increase their stellar masses via star formation and mergers. Star formation can be classified as quiescent star formation and starburst. Quiescent star formation happens in non-merger galaxies with typical star formation rate less than a few solar masses per year, while starburst is triggered by gas-rich mergers with star formation rate up to thousands of solar masses per year \citep[e.g.][]{Mihos1996, Springel2000, Tissera2002, Meza2003, Cox2006, Matteo2008}. 
 \citet{Guo2008} found in a semi-analytical galaxy catalogue that star formation dominates the growth of low mass galaxies across the cosmic time, while mergers are dominant for massive galaxies at low redshifts. Similar results have been reported by \citet{Qu2017} using EAGLE hydrodynamic simulations. Observationally, taking advantage of the Hubble Space Telescope (HST) images, \citet{Kaviraj2013} argued that galaxies mainly grow through ``secular processes" rather than by merging with other galaxies. Using the small-scale two point correlation functions, \citet{Bell2006} claimed major mergers are important for massive galaxies at low redshifts. Although many efforts have been devoted to understanding the role star formation plays in galaxy growth, few have been concerned about the effect of starbursts.

In this study, we apply a semi-analytic galaxy formation model to investigate the relative contributions of mergers, radio mode gas accretion/quiescent star formation, and quasar mode gas accretion/starburst to the mass assembly of SMBHs and their host galaxies, with the hope of shedding more light on understanding the respective roles of these processes in SMBH formation and galaxy growth, and how they co-evolve over cosmic time. 

This paper is composed of the following parts. In Section \ref{sec_sim}, we briefly introduce the $N$-body simulation and semi-analytic model. In Section \ref{sec_res}, we investigate the relative roles of mergers, radio mode gas accretion/quiescent star formation, and quasar mode gas accretion/starbursts in the mass build-up of the SMBHs and their host galaxies. Our results are summarized in Section \ref{sec_con}.

\section{Simulations and semi-analytic galaxy formation model} \label{sec_sim}

We study the effects of different growth channels including mergers, radio mode gas accretion/quiescent star formation, and quasar mode gas accretion/starbursts for the SMBHs and galaxies using the L-Galaxies semi-analytical model  \citep[][here after Guo2013]{Guo2013}, which is the updated version of previous L-Galaxies semi-analytical model \citep[e.g.][]{Springel2005, Croton2006,DeLucia2007}.

The Guo2013 model was implemented onto the merger trees extracted from the Millennium-W7 $N$-body simulation\citep{Guo2013}, which follows $2160^3$ dark matter particles in a box of $500~h^{-1}\mathrm{Mpc}$ on each side. The dark matter particle mass is $9.36\times 10^8~h^{-1}\rm M_{\odot}$. It adopted the cosmological parameters from the seven-year Wilkinson Microwave Anisotropy Probe (WMAP7): $\Omega_{\rm m}=0.272$, $\Omega_\Lambda=0.728$, $\sigma_8=0.807$, $h=0.704$ and $n_{\rm s}=0.961$.   It follows dark matter particles from $z = 127$ to the present day. Haloes are identified utilizing a friends-of-friends (FoF) algorithm with a linking length of 0.2 times the mean particle separation \citep{Davis1985}, and the gravitational self-bound substructures (subhaloes) within the haloes are identified applying the SUBFIND algorithm \citep{Springel2001}. The merging history of haloes (merger trees) is built up employing the LHALOTREE algorithm \citep{Springel2005}.

 In Guo2013, a given galaxy is seeded in the center of a halo when it is first identified, and follows its host halo until it merges into another big system and gets disrupted due to tidal forces. At this point, the galaxy becomes an orphan galaxy and then either merges with the center galaxy or gets disrupted by environmental effects. The semi-analytical galaxy formation model implements a comprehensive set of prescriptions including star formation, stellar evolution and feedback, SMBH accretion and AGN feedback, etc. It has been proven successful in reproducing many observed galaxy properties, both in the local Universe and at high redshifts, including the stellar mass functions, luminosity functions, stellar mass vs. metallicity relations, color distribution, and correlation functions \citep[e.g.][]{Guo2013,Marulli2013,Xie2014} . We refer readers to the papers introducing the projects for more details on the simulation setup and the galaxy formation prescriptions. Here we briefly describe the relevant processes.

In Guo2013, it is assumed that at high masses gas falls into gravitationally dominant dark matter haloes, shock-heated  to the virial temperature, and then cools and condenses into the center of the potential well. For low-mass systems, gas could reach the potential well as free-fall for their cooling time is very short. In both cases, gas ends up as a rotationally supported disk due to the conservation of angular momentum. When the local gas mass of the disk($M_{\rm gas}$) exceeds a critical value, $M_{\rm crit}$, gas in the disk will be converted into stars following:
 \begin{equation} \label{eq_sf}
\dot{M}_{*} = \alpha(M_{\rm gas}-M_{\rm crit})/t_{\rm dyn},
\end{equation}
where 
\begin{equation} \label{eq_mcrit}
M_{\rm crit}=11.5\times
10^9\left(\frac{V_{\rm max}}{200~{\rm km~s^{-1}}}\right)\left(\frac{R_{\rm disk,gas}}{10~{\rm kpc}}\right){\rm M}_{\odot}
\end{equation}
Here $R_{\rm disk,gas}$ is the exponential scale length of the gas disk, $t_{\rm dyn} = 3R_{\rm disk,gas}/V_{\rm max}$ is the dynamical timescale at the edge of the gas disk, $V_{\rm max}$ is the maximum circular velocity, and $\alpha=0.011$ is the star formation efficiency. 

We rebuild galaxy merging histories from the merger trees of their host dark matter haloes. The galaxy merger trees connect a galaxy to its progenitors at different redshifts, from which we can estimate the mass acquired through mergers. Merger events can trigger an intense burst of star formation if the progenitor galaxies are gas-rich.   
Adopting the burst model in \citet{Somerville2001}, the fraction of cold gas that is converted into stars during a merger-driven starburst is calculated as follows
 \begin{equation} \label{eq_burst}
e_{\rm burst} = 0.56\Big(\frac{M_{\rm sat}}{M_{\rm cen}}\Big)^{0.7},
\end{equation}
where $M_{\rm cen}$ and $M_{\rm sat}$ are the stellar masses of the central and the satellite galaxies, respectively. 

The SMBHs can grow via two modes of accretion: quasar mode and radio mode. The quasar-mode accretion is active only during mergers and the corresponding gas accretion is described as 
\begin{equation} \label{eq_bh_quasar-mode}
\delta M_{\rm BH} =  f\Big(\frac{M_{\rm sat}}{M_{\rm cen}}\Big)\Big[\frac{M_{\rm cold}}{1+(280~{\rm km~ s^{-1}}/V_{\rm vir})^2}\Big],
\end{equation}
where $f$ is a free parameter, set to be $f=0.03$. $M_{\rm cold}$ is the total mass of the cold gas in two merging galaxies and $V_{\rm vir}$ is the virial velocity of the central halo. The radio mode accretion occurs whenever the $M_{\rm BH}$ is non-zero and there is hot gas around the host galaxy
\begin{equation} \label{eq_bh_radio-mode}
\dot {M}_{\rm BH} = \kappa\Big(\frac{f_{\rm hot}}{0.1}\Big)\Big(\frac{V_{\rm vir}}{\rm 200~km~ s^{-1}}\Big)^3\Big(\frac{M_{\rm BH}}{10^{8}~h^{-1}{\rm M}_\odot}\Big){\rm ~M_\odot yr^{-1}},
\end{equation}
where $\kappa$ is a free parameter to describe the efficiency of hot gas accretion for which we adopt $\kappa=7\times10^{-6}$ in order to match the observed abundance of high-mass galaxies, and $f_{\rm hot}$ is the ratio of hot gas mass-to-halo mass. Previous works based on earlier versions of L-Galaxies \citep[e.g.][]{Croton2006,DeLucia2007} have been able to reproduce many observed AGN properties, including luminosity functions and correlation functions \citep[e.g.][]{Marulli2008,Bonoli2009}. In addition to merger induced quasar mode accretion, some other semi-analytic models \citep[e.g.][]{Bower2006,Marshall2020} introduce quasar mode accretion triggered by disk instabilities, which also reproduce many observed AGN properties. The disk instability could be triggered by interaction with small galaxies, and thus the SMBH growth by minor mergers and by disk instability could be at least partly degenerated. On the other hand, observationally, it is also not clear whether disk instability is related to quasar activities \citep[e.g.][]{Ho1997,Chen2013,Goulding2017,Galloway2015,Alonso2013,Kim2020}. Detailed analysis is beyond the scope of this paper.

\begin{figure} 
\centering\includegraphics[width=0.7\linewidth]{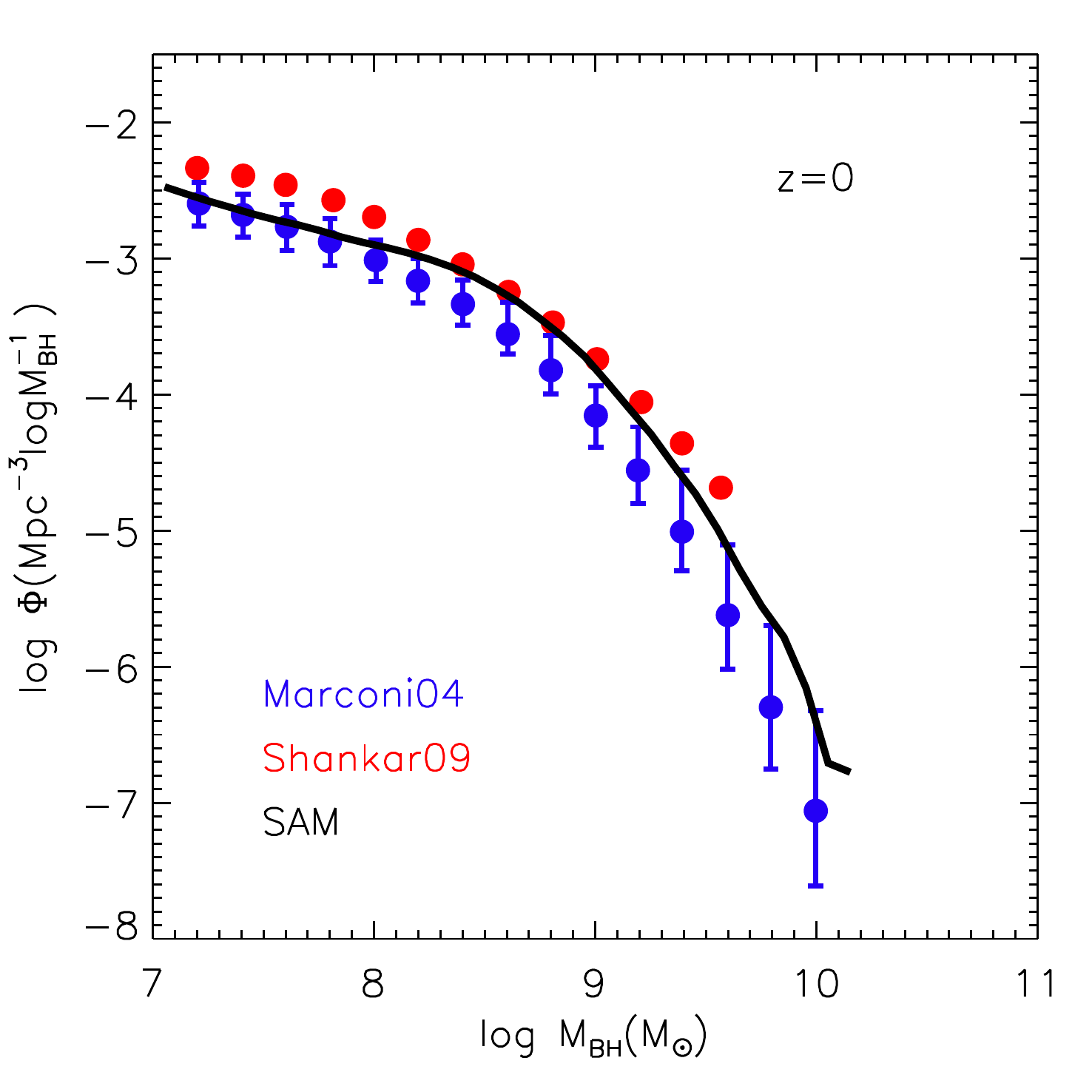} 
\caption{The SMBH mass functions at $z=0$ predicted by Guo2013 (black curve). Observational results are presented using blue circles \citep{Marconi2004} and red circles \citep{Shankar2009}.}\label{BHMF}
\end{figure}

In Fig. \ref{BHMF}, symbols represent the SMBH mass functions obtained by applying the scaling relation between SMBH mass and galaxy properties to the observed galaxy luminosity and velocity dispersion functions. These SMBH mass functions agree with that derived from AGN relics \citep[e.g.][]{Marconi2004,Shankar2009}. The predicted SMBH mass function is displayed with a black curve, which is consistent with the observational results in the mass range explored. 

\begin{figure} 
\centering\includegraphics[width=0.7\linewidth]{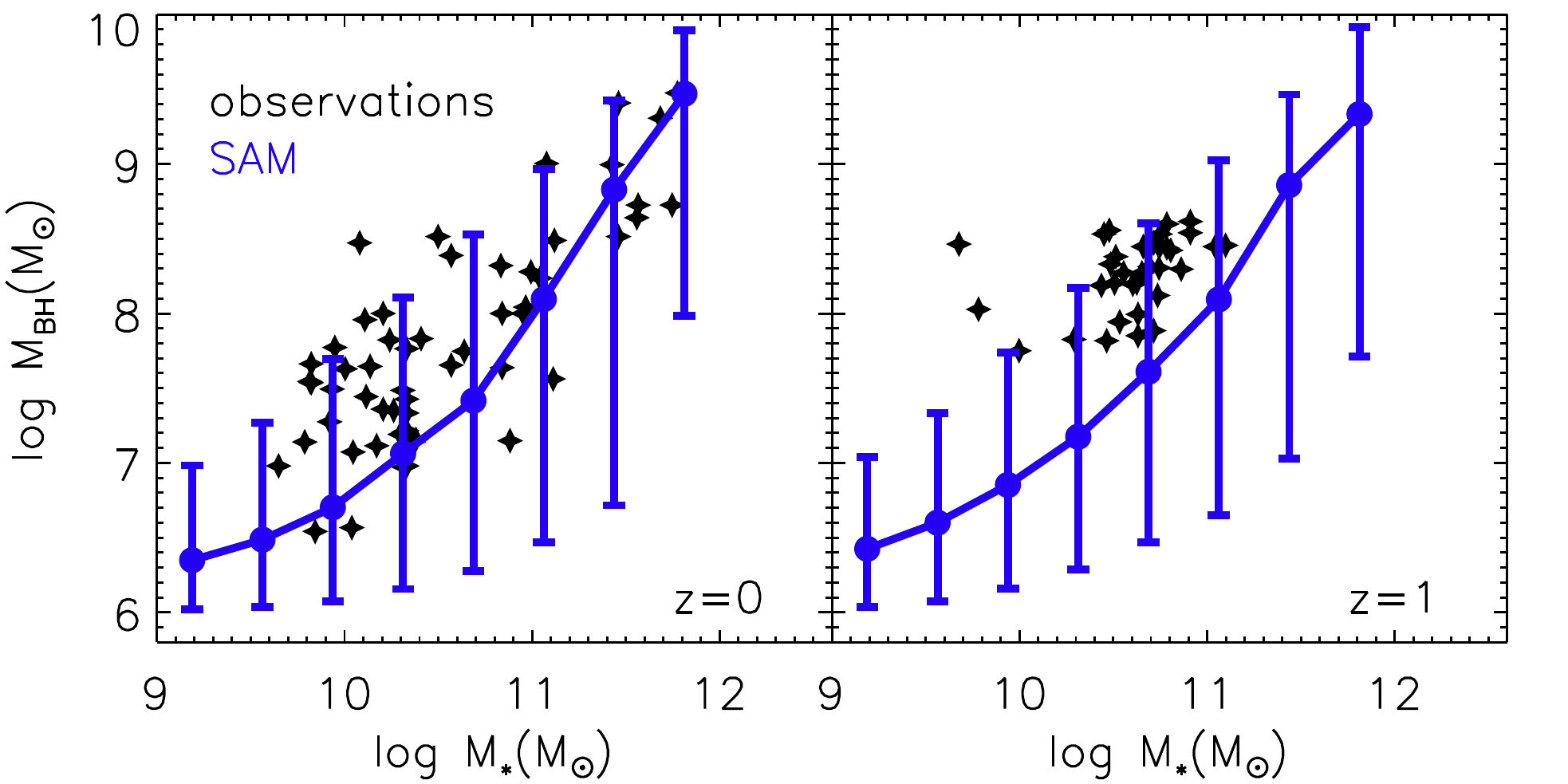}  
\caption{SMBH mass vs. the stellar mass relations. The predicted median values are presented using blue curves. Error bars present the 2$\sigma$ deviations from the median values. Black symbols are observational results in the local Universe  \citep{Bennert2011,Haring2004} and results at $z=1$ \citep{Ding2020,Ding2021}.}\label{Mbh-Mstar}
\end{figure}

The scaling relations between SMBH mass and stellar mass are regarded as evidences of the co-evolution between the SMBH and its host galaxy. Here we compare the predicted relations  with observations at different redshifts in Fig. \ref{Mbh-Mstar}. The predicted SMBH mass increases with the host galaxy stellar mass with scatters as large as three orders of magnitude at $z = 0$. Note the relations between the SMBH mass and the galaxy bulge stellar mass are tighter \citep[e.g.][]{Kormendy2013}, which is also found in hydrodynamical cosmological simulations  \citep[e.g.][]{Sijacki2015}. The slope gets slightly shallower at z = 1. Symbols in the left panel show the observational results in the local Universe \citep{Bennert2011,Haring2004}.  \citet{Ding2021,Ding2020} extend the measurements to $z\sim1$ using AGNs whose host galaxy images can be acquired by the Hubble Space Telescope. It demonstrates that the model predictions are consistent with the observational results, though the observed median values are slightly higher than the predicted ones.  The SMBH functions and $M_{\rm BH}$ vs. $M_{*}$ scaling relations are also reproduced in earlier versions of L-Galaxies \citep[e.g.][]{Croton2006,DeLucia2007,Marulli2008,Bonoli2009}, as well as in many other models and hydrodynamical simulations \citep[e.g.][]{Bower2006,Marulli2007,Fanidakis2012,Griffin2019,Marshall2020,Habouzit2021}.

\section{Results}
\label{sec_res}

\subsection{Mass growth of the SMBH}\label{result_1}

When two SMBHs with masses $M_1$ and $M_2$ ($M_1>M_2$) merge, we refer it to as a major merger event if $M_2/M_1>1/3$, otherwise a minor merger. We follow \citet[]{Guo2008} (hereafter Guo08) to define a dimensionless major merger rate as
 \begin{equation} \label{eq_major_rate}
R(M_{\rm BH},z) = \frac{N_{\rm major}(M_{\rm BH}, z)/\delta{ t}}{N(M_{\rm BH}, z)/t(z)},
\end{equation}
where $N(M_{\rm BH}, z)$ is the number of SMBHs at redshift $z$ with mass $M_{\rm BH}$, $N_{\rm major}(M_{\rm BH}, z)$ is the number of SMBHs that just underwent a major merger at redshift $z$, $t(z)$ is the age of the Universe at redshift $z$ and $\delta {t}$ is the time interval between the snapshot at $z$ and the one just before $z$.

\begin{figure} 
\centering\includegraphics[width=0.7\linewidth]{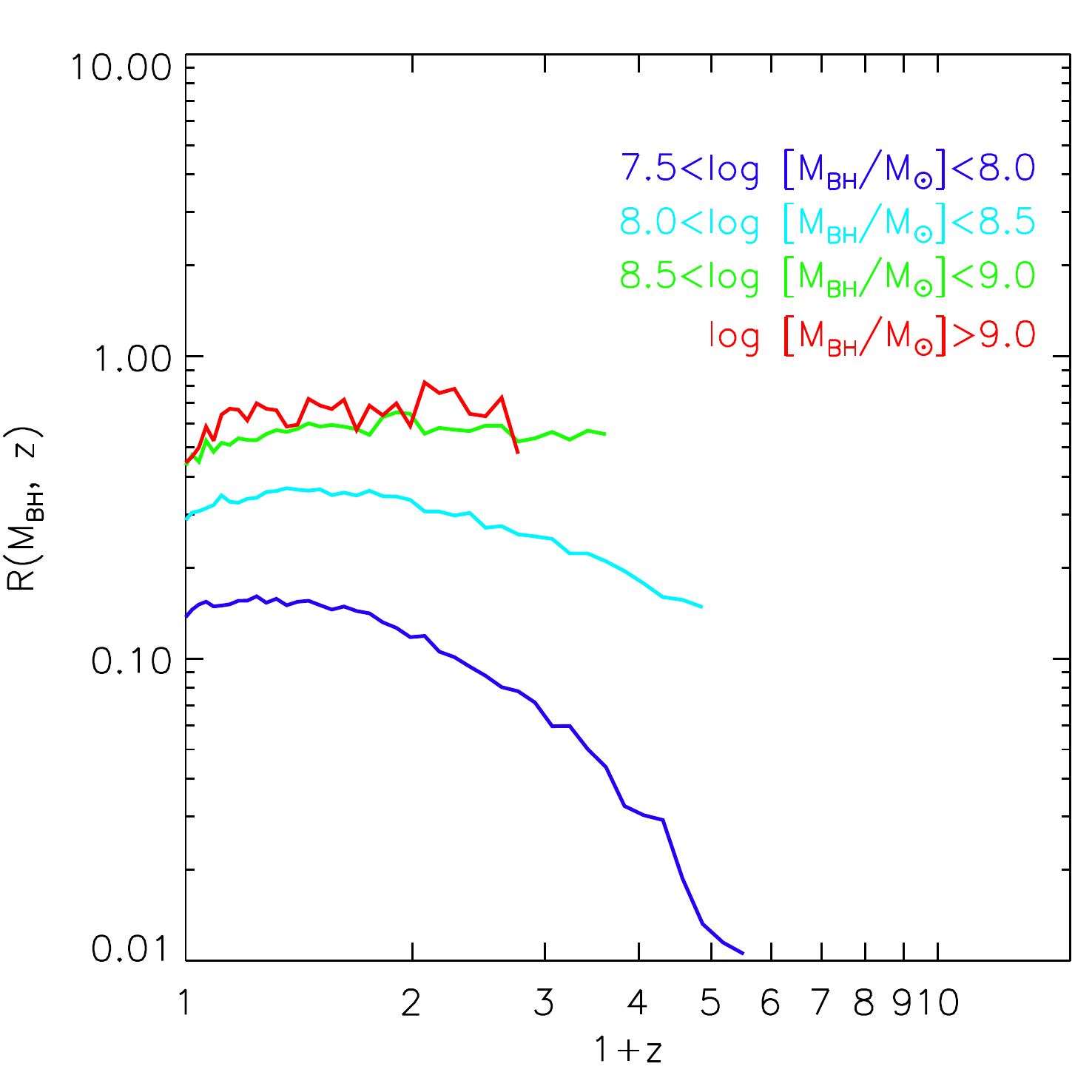}
\caption{SMBH major merger rates as a function of mass and redshift. The SMBHs are grouped into four mass bins, as denoted in the right upper corner. Results for SMBH of mass $10^{7.5}-10^{8}$ ${\rm M}_\odot$, $10^{8}-10^{8.5}$ ${\rm M}_\odot$, $10^{8.5}-10^{9}$ ${\rm M}_\odot$ and $>10^{9}$ ${\rm M}_\odot$ are presented using blue,cyan, green and red curves, respectively. 
 }\label{BH-Major-rate}
\end{figure}

We divide the SMBH samples into four mass ranges, $10^{7.5}-10^{8}$ ${\rm M}_\odot$ (blue), $10^{8}-10^{8.5}$ ${\rm M}_\odot$ (cyan), $10^{8.5}-10^{9}$ ${\rm M}_\odot$ (green) and $>10^{9}$ ${\rm M}_\odot$ (red). The dependence on masses and redshifts of the major merger rates are presented in Fig. \ref{BH-Major-rate}. The mass dependence and redshift dependence of  major merger rate vary with the SMBH masses. The massive SMBHs with $M_{\rm BH}>10^{8.5}~{\rm M}_\odot$ have the highest major merger rate. Their mass dependence is weak and there is almost no variation with redshift. The major merger rates drop dramatically towards lower masses. At $M_{\rm BH}<10^{8.5}~{\rm M}_\odot$ the major merger rates increase with redshift slightly and then decrease at high redshifts. Similar mass and redshift dependencies are found for galaxies in \citet{Guo2008}. 

We further define a dimensionless mass growth rate $R_{\rm m}$ for the SMBHs as 
 \begin{equation} \label{eq_bh_growth_rate}
R_{\rm m}(M_{\rm BH},z) = \frac{\delta M_X(M_{\rm BH}, z)/\delta{ t}}{M(M_{\rm BH}, z)/t(z)},
\end{equation}
where $M(M_{\rm BH}, z)$ is the total mass of the SMBHs at redshift $z$ with mass $M_{\rm BH}$, $\delta M_{\rm X}(M_{\rm BH}, z)$ is the total mass obtained via different mechanisms in the time interval $\delta t$, X stands for the mass obtained via major mergers, all mergers, total gas accretion and quasar mode gas accretion.
The contribution from all mergers is the sum of the mass of the swallowed SMBHs. Removing this mass from the total SMBH mass growth leaves us the mass that SMBHs obtain via total gas accretion, $\delta M_{\rm acc}(M_{\rm BH}, z)$. In SAM, the mass growth via quasar mode accretion, $\delta M_{\rm quasar}(M_{\rm BH}, z)$, is recorded. The mass growth via radio mode, $\delta M_{\rm radio}(M_{\rm BH}, z)$, is calculated as $\delta M_{\rm acc}- \delta M_{\rm quasar}$. 

\begin{figure*} 
\centering\includegraphics[width=\linewidth]{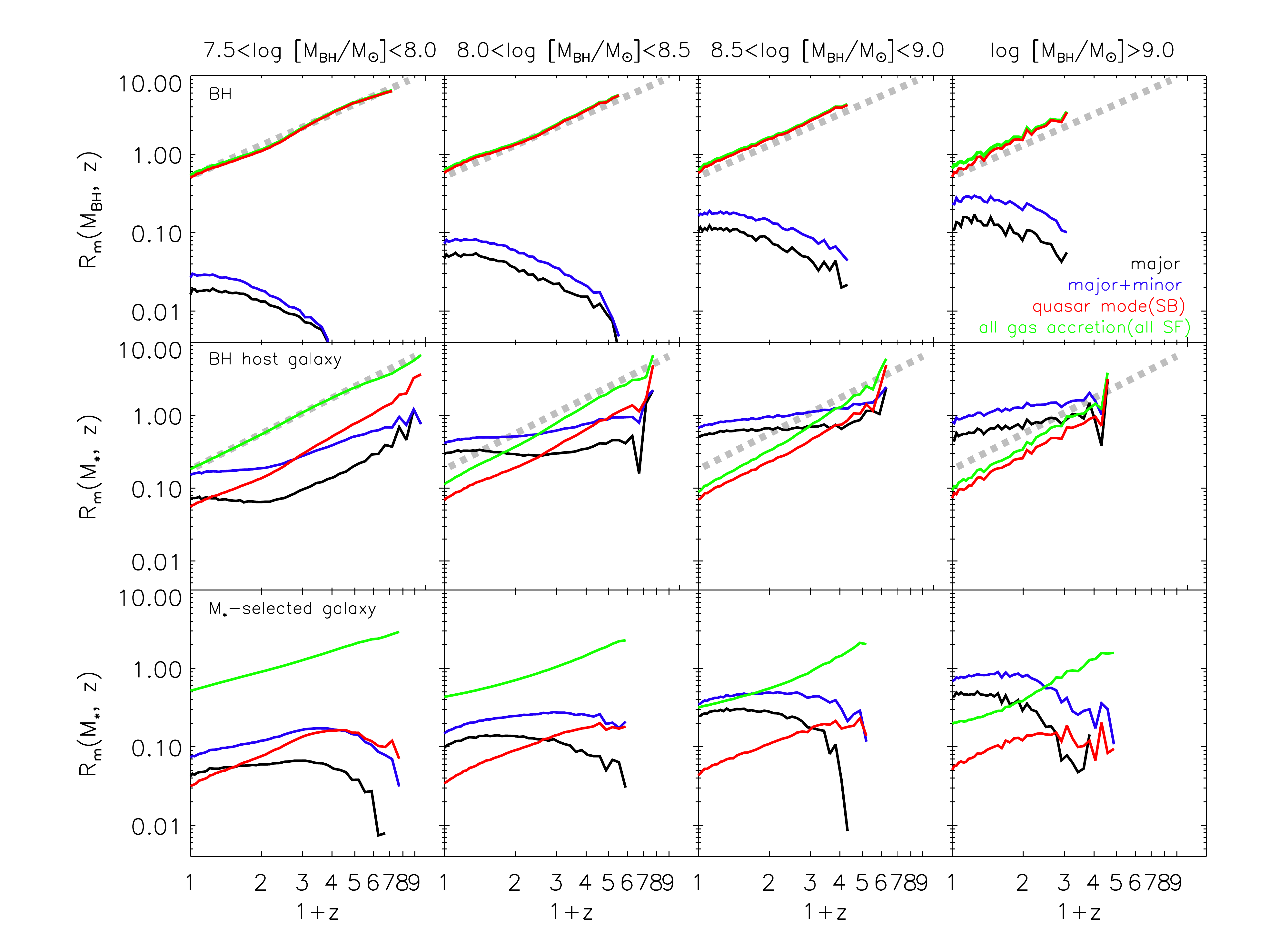} 
\caption{Dimensionless mass growth rates as a function of redshifts. It displays results for the SMBHs, for the corresponding SMBH host galaxies, and for galaxies selected by stellar masses from the top to the bottom rows, respectively. Different columns are for different mass ranges. Growth rates by mergers, major mergers, gas accretion/star formation, and quasar mode accretion/starburst are presented using blue, black, green and red curves, respectively. Dashed grey lines in the top row are the power law fits to the growth rate via gas accretion in the top left panel, while dashed grey lines in the middle row are the power-law fits to the growth rate via star formation in the middle left panel.}\label{BH-growth-rate}
\end{figure*}

The first row of Fig. \ref{BH-growth-rate} depicts the dimensionless mass growth rate $R_{\rm m}(M_{\rm BH},z)$ as a function of SMBH mass and redshift. We find clearly that the gas accretion is the driving mechanism of the SMBH growth throughout the redshift and mass ranges we explored. Its contribution is orders of magnitude higher than mergers at high redshifts. At low redshifts, it is still several times higher. The relative importance from mergers increases with increasing SMBH mass but remains sub-dominant at all masses and at all redshifts. At high masses, mergers start to play a more important role, while at low masses, the contribution from mergers is always negligible. These relative changes are consistent with the increasing major merger rate with the SMBH mass. 

Previous works mostly focus on the accumulative contribution to the SMBH growth. \citet{McAlpine2020} found in EAGLE that the excess of merger fractions between active and inactive
galaxies is restricted to galaxies less massive than $\rm 10^{11}~{\rm M}_{\odot}$, while \citet{Steinborn2018} found in the cosmological hydrodynamical Magneticum Pathfinder simulations that the merger fractions of AGN hosts can be up to three
times higher than those of inactive galaxies at $M_*>10^{11}~{\rm M}_{\odot}$. Despite the differences, they both find that galaxy mergers are not relevant in the SMBH growth. \cite{McAlpine2020} also found the accumulative contribution from major mergers increases with the final SMBH mass, up to 40\% at $M_{\rm BH}(z=0)= 10^9~{\rm M}_{\odot}$. \citet{Marshall2020} observed in the Meraxes model that gas accretion dominates the SMBH growth. This is consistent with our results. However, \citet{Fanidakis2011} and \citet{Dubois2014} found in the GALFORM semi-analytical models \citep{Bower2006} and in a cosmological hydrodynamical simulation that the dominant channel of SMBH growth is the quasar mode accretion only for SMBHs less massive than $10^8~{\rm M}_{\odot}$, while at high masses and at low redshifts mergers can be the dominant growth channels, contradictory to our findings. Similar results are ascertained by \citet{Weinberger2018} in IllustrisTNG that the instantaneous growth of the SMBH is dominated by gas accretion at all times for low mass SMBHs, while mergers become entirely dominant at high masses at low redshifts. They argue that it could be due to the different feedback efficiencies. 

 The top left panel indicates that the SMBH mass growth rate via gas accretion increases with redshifts, roughly as a power law of $(1+z)^{1.32}$, indicating a faster growth of the SMBH at higher redshifts.  
 The quasar mode accretion always dominates over the radio mode accretion, similar to the conclusion given by \citet{Croton2006} and \citet{Marshall2020}. The relative role of radio mode accretion compared to the quasar mode accretion increases with SMBH mass but is never in excess of unity. This is in contrast with that found by  \citet{Fanidakis2011} and \citet{Griffin2019} in GALFORM that the contribution of radio mode accretion could exceeds quasar mode accretion for massive SMBHs at low redshifts.
 
SMBH mass growth rate via mergers decreases with increasing redshifts at $z>0.3$, suggesting a slower growth of the SMBH in the earlier Universe. We further separate the mass growth via major mergers from minor mergers. It shows that the major mergers are more important compared to minor mergers at low masses, while at high masses, the contributions from major and minor mergers are comparable.

\subsection{The stellar mass assembly of SMBH host galaxies}\label{result_2}
In this section we investigate the stellar mass growth histories of the SMBH host galaxies and compare to the stellar mass-selected galaxies. We replace $M_{\rm BH}$ in equation (\ref{eq_bh_growth_rate}) with  $M_*$  to obtain the galaxy growth history via each mechanism. Since the quasar mode accretion and the starburst are both triggered by mergers, we thus compare their contributions to the SMBH growth and to the galaxy growth. Similarly, we compare the SMBH growth via gas accretion to the galaxy growth via star formation.

The middle row of Fig. \ref{BH-growth-rate} displays the results for the SMBH host galaxies. Mergers are very important in their growth histories. For galaxies which host the most massive SMBHs, mergers are the dominant mechanism up to $z\sim3$. At lower masses, mergers are still dominant at low redshifts. Only for galaxies hosting the least massive SMBHs does the star formation dominate their growth at all epochs. The redshift at which mergers start to take over increases with their SMBH mass. The mass growth rate via mergers increases by about an order of magnitude from the lowest mass range to the highest one. In general, mergers play a much more important role in the growth of the SMBH host galaxies than in the growth of the SMBHs themselves. 

The stellar mass growth rate by star formation increases monotonically with redshift as a power law but with a steeper slope,  $\sim(1+z)^{1.63}$, compared to the SMBH growth rate via gas accretion. This steeper slope suggests a flattening in the $M_{\rm BH}$ vs. $M_*$ relation towards high redshifts, consistent with what we find in Fig.2.

Despite the fact that both starburst and quasar mode accretion are triggered by mergers, their relative roles in galaxy and SMBH growth are quite different. 
The quasar mode accretion plays a dominant role in the SMBH growth over all masses, while starbursts dominate over the quiescent star formation only at high masses for the host galaxies. This is consistent with the more and more important roles that mergers play with increasing masses. 

Interestingly, the relative contributions to the galaxy stellar mass growth via starburst and mergers vary with mass and redshifts, though the starburst is triggered by mergers by construction. The contribution from starburst increases with redshift as a power law, while the contribution from mergers varies very little with redshifts. These are  related to the fact that at high redshifts mergers are usually gas-rich and involve more extensive star formation as a consequence.  

Are these SMBH host galaxies special? In the bottom row in Fig. \ref{BH-growth-rate} we show the contributions of different growth modes for galaxies selected by stellar mass. We split galaxies into four stellar mass bins according to the SMBH mass ranges. In practice, we use the median value of $M_*$ for a given $M_{\rm BH}$ in Fig. \ref{Mbh-Mstar}. The corresponding stellar mass bins are $10^{10.33}-10^{10.64}$ ${\rm M}_\odot$, $10^{10.64}-10^{10.96}$ ${\rm M}_\odot$, $10^{10.96}-10^{11.27}$ ${\rm M}_\odot$, $>10^{11.27}$ ${\rm M}_\odot$. We find that star formation dominates the assembly of the stellar mass selected galaxies at all epochs below $10^{10.96}~{\rm M}_\odot$. The relative importance of mergers increases with stellar mass, and finally mergers take over as the dominant mechanism for massive galaxies at low redshifts. At high redshifts, star formation is always the the predominant channel. This is consistent with those found in previous studies \citep[e.g.][]{Guo2008, Qu2017}. 

Different from the SMBH host galaxies,  the slope of the growth rate via star formation is flatter and the contribution from starburst is always negligible compared to the contribution from quiescent star formation. 

We summarize that gas accretion dominates the growth of the SMBHs, while mergers are more important in the growth of their host galaxies, especially at high masses. The SMBH growth and the host galaxy growth do not follow each other in a simple manner. Mergers are more important for the growth of the SMBH host galaxies than for normal galaxies.

\begin{figure*}[h]
\centering\includegraphics[width=\linewidth]{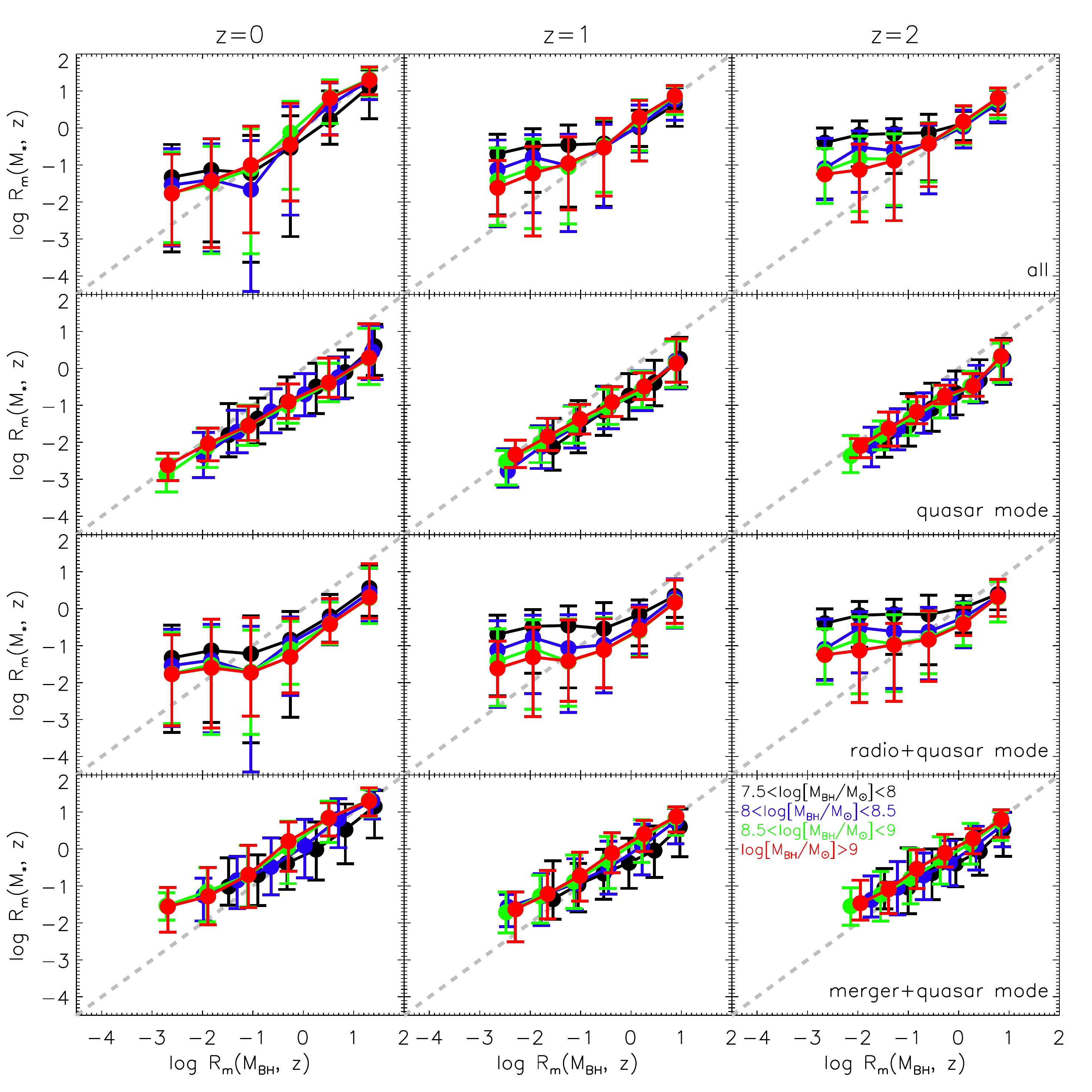} 
\caption{Growth rates of SMBHs vs. growth rates of their host galaxies. Different rows correspond to results from different mechanisms. The first row is the total growth rates, the second row is the growth rates via quasar mode/starburst, the third row is the combined accretion rates via quasar mode and radio mode accretion compared to the combined stellar mass growth rates via starburst and quiescent star formation, and the bottom row is the combined SMBH growth rates via mergers and quasar mode accretion compared to the combined host galaxy growth rates via mergers and starbursts. Different columns are for different redshifts as indicated on the top of each column. Different colors are for different SMBH mass ranges, as indicated in the last panel. Dashed grey lines signify the 1:1 growth rates. Error bars present the 2$\sigma$ deviations from the median values.
}\label{host-galaxy-growth-rate}
\end{figure*}

\subsection{The co-evolution between SMBHs and their host galaxies}
In this section, we compare the impact of gas accretion/star formation, mergers, and quasar mode accretion/starburst on the growth of the SMBHs and their host galaxies in a one-by-one manner, in order to shed more light on the co-evolution between the SMBHs and their host galaxies. 

The first row in Fig.\ref{host-galaxy-growth-rate} features the total SMBH mass growth rate vs. the total stellar growth rate of their host galaxies. It demonstrates that at $z=0$ when the mass growth rate exceeds 0.1, the growth of the SMBHs follows their host galaxies. At lower growth rate, the connection between the growth rate of the SMBH and their host galaxies disappears, and the galaxies tend to grow more efficiently than the SMBHs. Similar relations are found at higher redshifts. The characteristic growth rate above which the SMBHs growth follows their host galaxies increases with redshifts up to $R_{\rm m} \sim1$ at $z=2$.
We note that at high redshifts below the characteristic growth rate the dependence on SMBH mass affirms that galaxies hosting higher mass SMBHs have lower growth rates. 

We further separate the contributions from different growth mechanisms. In the second row we explore the SMBH growth rate via quasar mode accretion compared to the growth rate via starburst for their host galaxies. It shows that these two quantities correlate with each other with small scatters. There is almost no redshift or SMBH mass dependence. However the slope somehow deviates from the 1:1 line signifying that SMBH grows faster than their host galaxies at high $R_{\rm m}$. 

This is in broad consistence with what was found in the literature in that the SMBH accretion rates correlate with the star formation rates in their host galaxies. Yet the observational results are not the same in details.  \citet{Diamond-Stanic2012,Xu2015,Dong2016} concluded the slope between the SFR and the SMBH growth is shallower than 1, similar to what we found. However, \citet{Chen2013} and \citet{Yang2019} ascertained the slope is close to 1. \citet{Delvecchio2015} instead identified an increasing slope with redshifts. These differences could be partly due to sample selections. 

Though quasar mode accretion dominates the SMBH growths, we have to include the radio mode accretion and correspondingly the quiescent star formation to understand the total contribution from gas accretion and star formation for the SMBHs and their host galaxies, respectively. The third row in Fig.~\ref{host-galaxy-growth-rate} shows that this combined effect brings the slope close to 1 at high $R_{\rm m}$, though the amplitude is slightly lower, suggesting SMBH accretion is more efficient than star formation in their host galaxy. At low $R_{\rm m}$, on the other hand, the growth rate of the SMBH host galaxies is independent of the growth rate of the SMBHs, similar to the total growth rate. This is consistent with the fact that quiescent star formation plays a more important role in the growth of the host galaxies compared to the radio mode accretion in SMBHs' growth, as described in the last subsection. There is also a characteristic growth rate below which the correlation between the growth of the SMBHs and their host galaxies disappears. This characteristic growth rate increases with redshifts. In general it makes the co-growth of SMBHs and their host galaxies weaker when including the radio-mode accretion and quiescent star formation. 

Since mergers can be the driving mechanism for the growth of the SMBH host galaxies, in the last row we combine the contribution from the quasar mode accretion/starburst and the mergers. It demonstrates clearly that the combined growth rates follow each other tightly along the 1:1 line at all growth rates and at all redshifts. In addition, we do not find any dependence on SMBH mass or redshifts. 

In sum it is the combined contribution from the quasar mode accretion/starburst and mergers that drives the co-evolution of the SMBHs and their host galaxies.

\section{Conclusions}\label{sec_con}

In this work, we investigate the mass assembly histories of the SMBHs and their host galaxies in simulated galaxy catalogues. We focus on several mechanisms through which the SMBHs and their host galaxies grow in mass, including gas accretion/star formation, quasar mode accretion/starburst and mergers. Our main results are summarized as follows:

\begin{itemize}

\item It is the combinations of mergers and quasar mode accretion/starburst that drive the co-evolution of the SMBH and their host galaxies.

\item It is the gas accretion that dominates the mass growth of the SMBH at all the mass ranges and redshift ranges explored. Mergers only start to play a significant role at low redshift at high masses but never dominate. Quasar mode accretion is more important than the radio mode accretion. 

\item Mergers play a much more important role in the growth of the SMBH host galaxies than in the growth of the SMBHs themselves. They can be the main contributor for galaxies which host SMBHs more massive than $10^8~{\rm M_{\odot}}$, especially at low redshifts. 

\item The SMBH host galaxies have different growth histories compared to normal galaxies. Mergers and starburst play more important roles in the assembly of the SMBH host galaxies.

\item The major merger rate of the SMBH below $10^{8.5}~{\rm M}_\odot$ is a strong function of mass, and decreases with increasing redshift at high redshifts. At higher masses, the mass and redshift dependencies are both weak. 

\end{itemize}

The hydrodynamical simulations have the advantages in tracing gas dynamics. Yet radio mode and quasar mode accretion are not well separated in such simulations. We plan to use the hydrodynamical simulations to further explore how mergers and the induced star formation (starburst) shape the growth of the SMBH in more detail in the future. This might shed more light on understanding the co-evolution between the SMBHs and their host galaxies. 

\normalem
\begin{acknowledgements}
  We thank referees for insightful suggestions that improved the manuscript. We thank Shihong Liao for discussions. Q.G. is supported by NSFC grants (Nos. 11573033, 11622325, 11425312 and 11988101), and the "Recruitment Program of Global Youth Experts" of China, the NAOC grant (Y434011V01). YQ is supported by NSFC grant (No. 11803045). L.G. is supported by the National Key R\&D Program of China (NO. 2017YFB0203300), and the Key Program of NFSC through grant 11733010. 
\end{acknowledgements}
  
\bibliographystyle{raa}
\bibliography{2021-0116}

\end{document}